\documentclass[final,1p,times]{elsarticle} 
\usepackage{graphicx} 
\usepackage{amssymb} 
\usepackage{amsthm} 
\usepackage{lineno} 
\usepackage{amsmath}

\def\third{\frac{1}{3}}
\def\twothirds{\frac{2}{3}}

\def\GeV{\mathrm{GeV}}

\journal{Nuclear Physics A} 
\begin{document} 

\begin{frontmatter} 


\title{Higher moments of charge fluctuations in QCD at high temperature}

\author{C.~Miao for the RBC-Bielefeld collaboration}

\address{}

\begin{abstract} 

We present lattice results for baryon number, strangeness and electric charge fluctuations as well as their correlations at finite temperature and vanishing chemical potentials, i.e. under conditions relevant for RHIC and LHC. We find that the fluctuations change rapidly at the
transition temperature $T_c$ and approach the ideal quark gas limit already at approximately $1.5T_c$. This indicates that quarks are the relevant degrees of freedom that carry the quantum numbers of conserved charges at $T\geq 1.5T_c$. At low temperature, qualitative features of the lattice results are well described by a hadron resonance gas model.

\end{abstract} 

\end{frontmatter} 



\section{Introduction}

Fluctuations of conserved charges, like baryon number, electric charge and strangeness, are generally considered to be sensitive indicators for the structure of a thermal medium produced in heavy ion collisions \cite{Koch}.
Under conditions met in current experiments at RHIC as well as in the upcoming heavy ion experiments at LHC the net baryon number is small and QCD at vanishing chemical potential provides a good approximation.
In this region the transition from the low temperature hadronic to the high temperature plasma regime is continuous and fluctuations are not expected to lead to any singular behavior. Indications for the existence of critical points can only show up in higher order derivatives of the QCD partition function with respect to temperature or chemical potentials \cite{Stephanov}. Through the analysis of fluctuations of conserved charges as well as their higher moments and correlations we thus gain insight into the relevant degrees of freedom of the system under consideration and at the same time gather information on possible nearby singularities in the QCD phase diagram.


At vanishing baryon number ($B$), electric charge ($Q$)
and strangeness ($S$) fluctuations of these quantities can
be obtained by starting from the QCD partition function
with non-zero light and strange quark chemical potentials,
$\hat\mu_{u,d,s} \equiv \mu_{u,d,s}/T$. The quark chemical
potential can be expressed in terms of chemical potentials
for baryon number ($\mu_B$), strangeness ($\mu_S$) and
electric charge ($\mu_Q$),
\begin{linenomath}
\begin{equation}
\mu_u = \third\mu_B + \twothirds\mu_Q\ ,\qquad
\mu_d = \third\mu_B - \third\mu_Q\ ,\qquad
\mu_s = \third\mu_B - \third\mu_Q - \mu_S\ .
\end{equation}
\end{linenomath}
Moments of charge fluctuations,
$\delta N_X \equiv N_X - \left< N_X \right>$,
with $X=B$, $Q$ or $S$ and their correlations are then obtained from derivatives of the logarithm of the QCD partition function, \emph{i.e.} the pressure,
evaluated at $\mu_{B,Q,S}=0$,
\begin{linenomath}\begin{equation}
\chi_{i,j,k}^{B,Q,S} = \left.\frac{\partial^{i+j+k}p/T^4}
    {\partial \hat\mu_B^i \partial \hat\mu_Q^j \partial \hat\mu_S^k}
    \right|_{\mu=0} \ ,
\end{equation}\end{linenomath} with $\hat\mu_X \equiv \mu_X/T$.
While the first derivatives, \emph{i.e.} baryon number, electric charge and strangeness densities, vanish for $B,Q,S = 0$, their moments and correlation functions with $i + j + k$ even are non-zero. The basic quantities we will analyze here are the quadratic, quartic and 6th order cumulant of fluctuations,
\begin{linenomath}
\begin{equation}
\chi_2^X = \frac{1}{VT^3} \langle N_X^2 \rangle,\quad
\chi_4^X = \frac{1}{VT^3} (\langle N_X^4\rangle
            - 3\langle N_X^2\rangle^2),\quad
\chi_6^X = \frac{1}{VT^3}(\langle N_X^6\rangle - 15\langle N_X^4\rangle\langle N_X^2\rangle +30\langle N_X^2\rangle^3).
\end{equation}
\end{linenomath}

\begin{figure}[t]
\includegraphics[width=.32\textwidth]{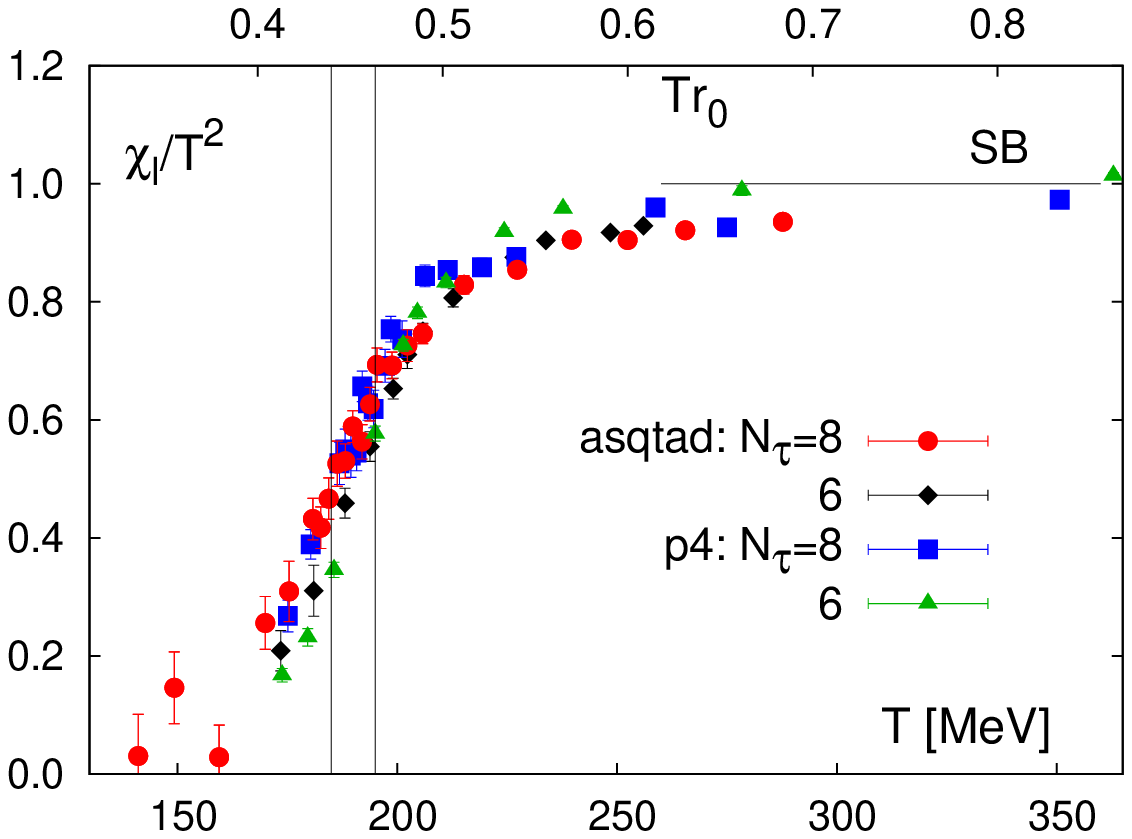}
\includegraphics[width=.32\textwidth]{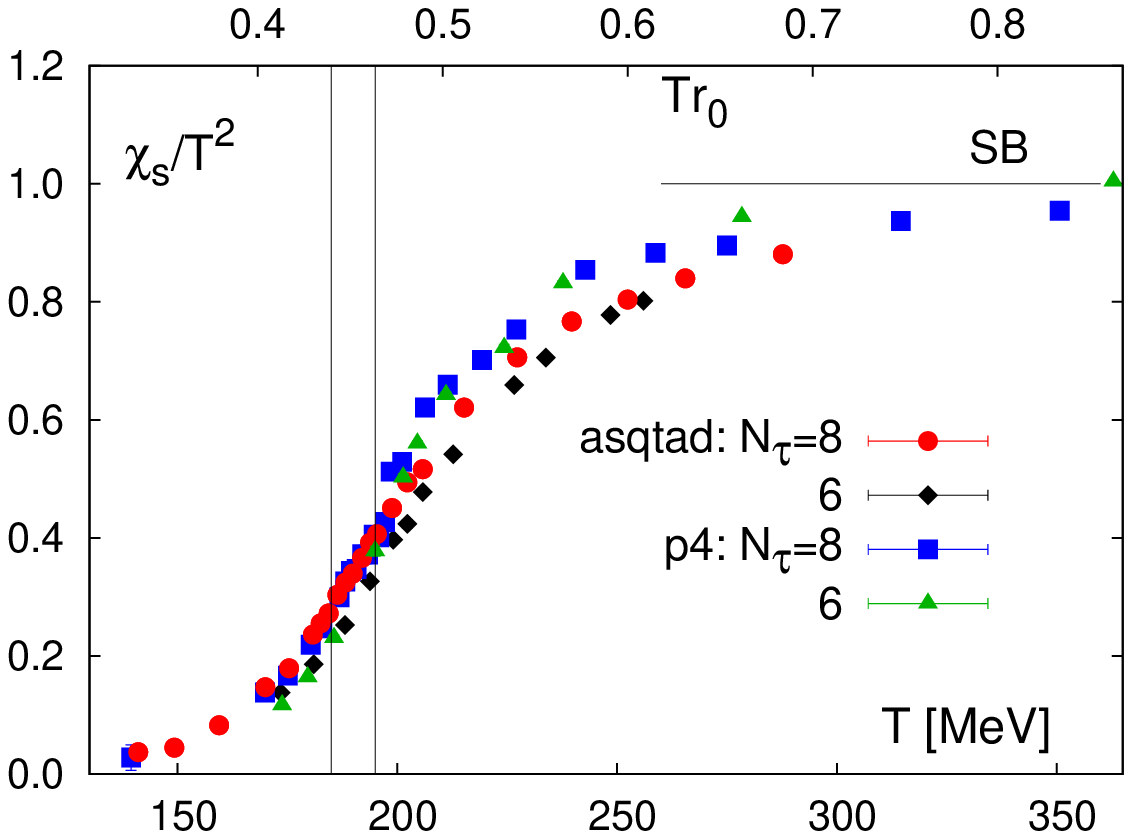}
\includegraphics[width=.32\textwidth]{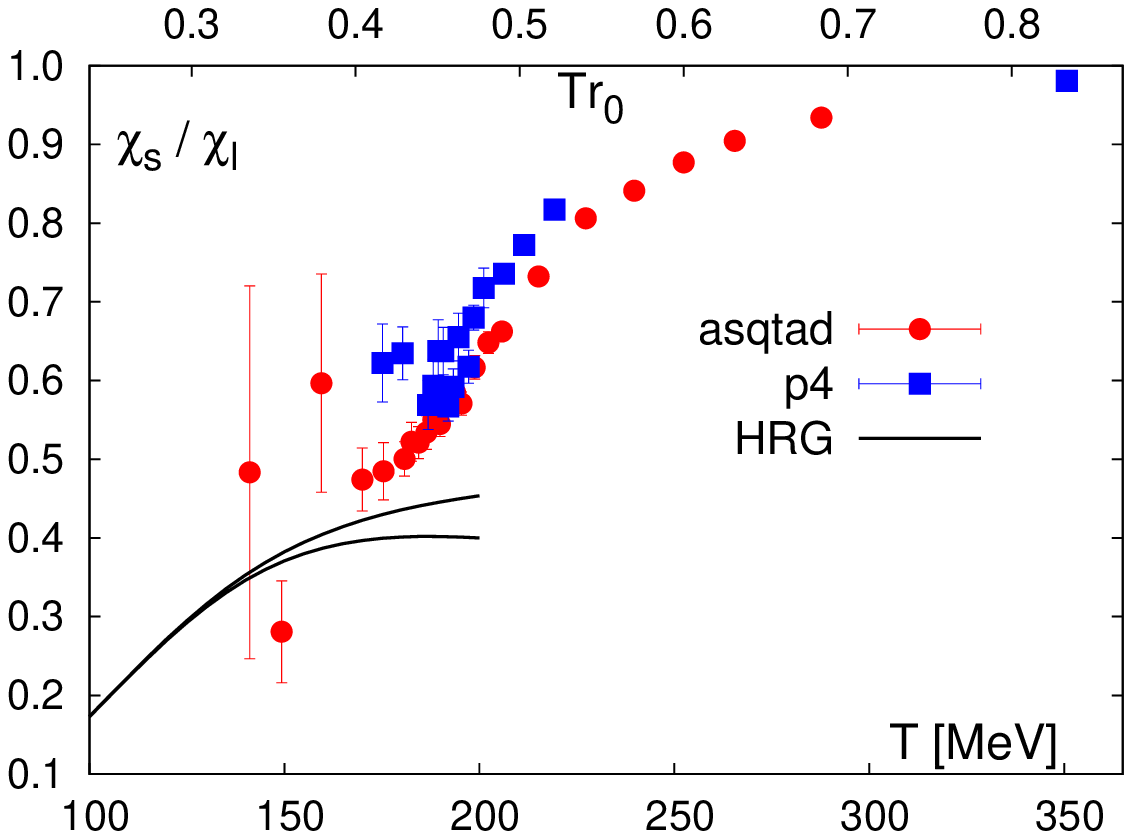}
\caption{The quadratic fluctuations of light (left) and strange (middle) quark number versus temperature on lattices of size $24^3\times6$ and $32^3\times8$. The horizontal solid lines show results for the ideal massless quark gas; the vertical lines mark the temperature interval $185~\mathrm{MeV}\leq T\leq195~\mathrm{MeV}$. The ratio of s and u quark fluctuations (right) are calculated on lattices with temporal extent $N_\tau=8$. Curves in the right hand figure show results for a hadron resonance gas including resonance up to $m_{max} = 1.5 \GeV$ (upper branch) and $2.5 \GeV$ (lower branch), respectively.}
\label{chi2}
\end{figure}

The gauge field configurations that have been used to evaluate the above observables, had been generated previously in calculations of the QCD equation of state \cite{ourEoS,hotEoS} and the transition temperature \cite{our_Tc,hotTc}. An improved staggered fermion action (p4-action) \cite{p4} that strongly reduces lattice cut-off effects in bulk thermodynamics at high temperature has been used. The strange quark mass has been tuned close to its physical value and the light quark masses have been chosen to be one tenth of the strange quark mass. This corresponds to a line of constant physics on which the kaon mass is close to its physical value and the lightest pseudo-scalar mass is about $220$ MeV.

These calculations have been performed on $16^3 \times 4$, $24^3 \times 6$ and $32^3\times8$ lattices. This allows us to judge the magnitude of systematic effects arising from discretization errors in our improved action calculations. The spatial volume has been chosen to be $V^{1/3}T = 4$, which insures that finite volume effects are small. In this calculations, the above observables have been evaluated in the temperature interval $0.8 \lesssim T/T_c \lesssim 2.5$.

\section{Results}
Before entering a discussion of fluctuations of B, Q and S it is instructive to look into fluctuations of the partonic degrees of freedom, the light and strange quarks. In Fig. \ref{chi2} we show the temperature dependence of quadratic fluctuations for light quarks $\chi_2^l$ and strange quarks $\chi_2^s$. They change rapidly in the transition region as the carriers of the quantum numbers are heavy hadrons at low temperatures but much lighter quarks at high temperatures. In the continuum and high temperature limit, these fluctuations  quickly approach the Stephan-Boltzmann limit, \emph{i.e.} an ideal massless quark gas.

The lattice cut-off effects for quadratic fluctuations are shown to be small by comparing the measurements on $N_\tau = 6$ and 8 lattices. Moreover, we have used two different cut-off schemes, p4 and asqtad \cite{asqtad}. Both schemes are improved up to $O(a^2)$ for thermodynamic quantities at high temperature. Indeed, the quark fluctuations have small deviations from SB limit at high temperature on $N_\tau=8$ lattices for both schemes. The measurements using p4 and asqtad actions are generally in good agreement, although we observe differences in the temperature interval between $T_c$ and $1.5~T_c$.

We observe that the strange quark fluctuations rise slower than the light quark fluctuations. As temperature decreases, their ratio  $\chi_2^s/\chi_2^l$ (Fig. \ref{chi2}) drops quickly in the transition region.The hadron resonance gas (HRG) model calculations show a similar trend of exponential fall at low temperature. This can be understood in the zero temperature limit: the light quark fluctuations are sensitive to pions, $\chi_2^l/T^2 \sim \exp(-m_\pi/T)$, while the strange quark fluctuations are sensitive to the lightest hadronic state that carry strangeness, $\chi_2^s/T^2 \sim \exp(-m_K/T)$. The HRG models can not reproduce the lattice measurements at higher temperature, as it breaks down at $T\simeq 150$~MeV, where HRG models with different spectrum cuts start to show deviations.

\begin{figure}[t]
\includegraphics[width=.32\textwidth]{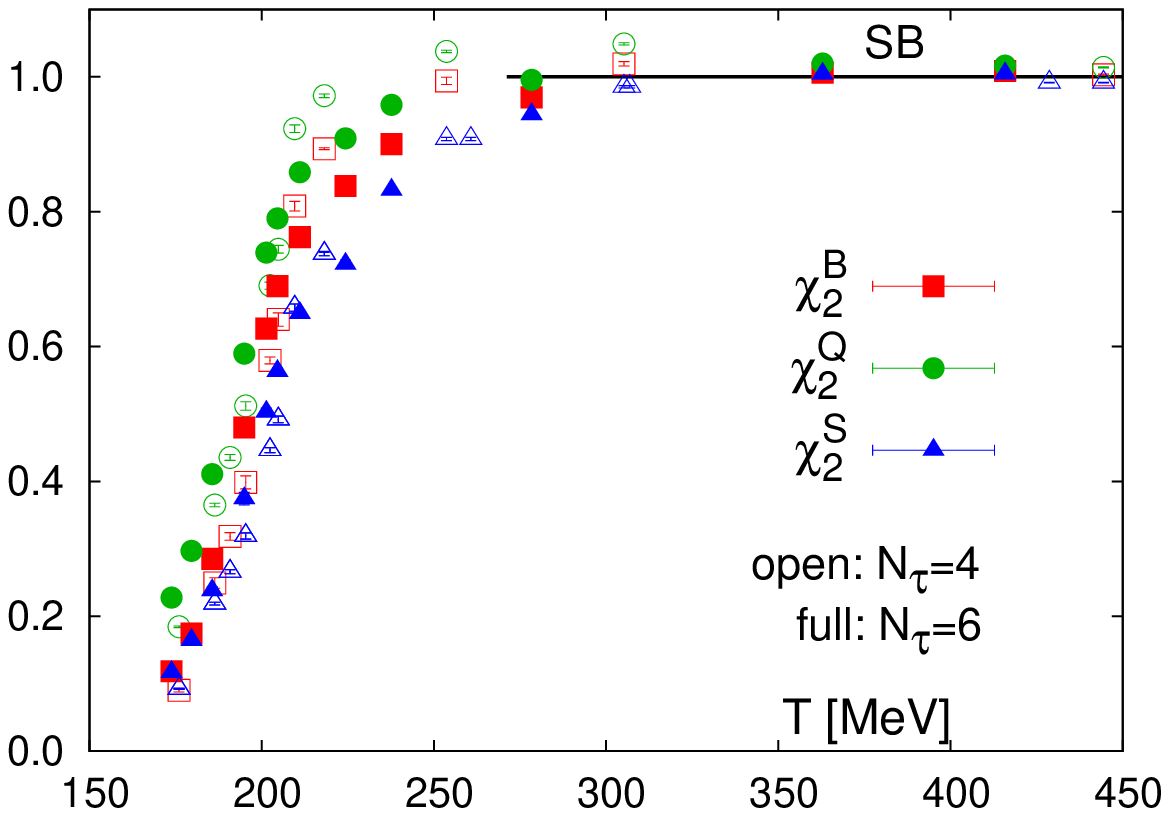}
\includegraphics[width=.32\textwidth]{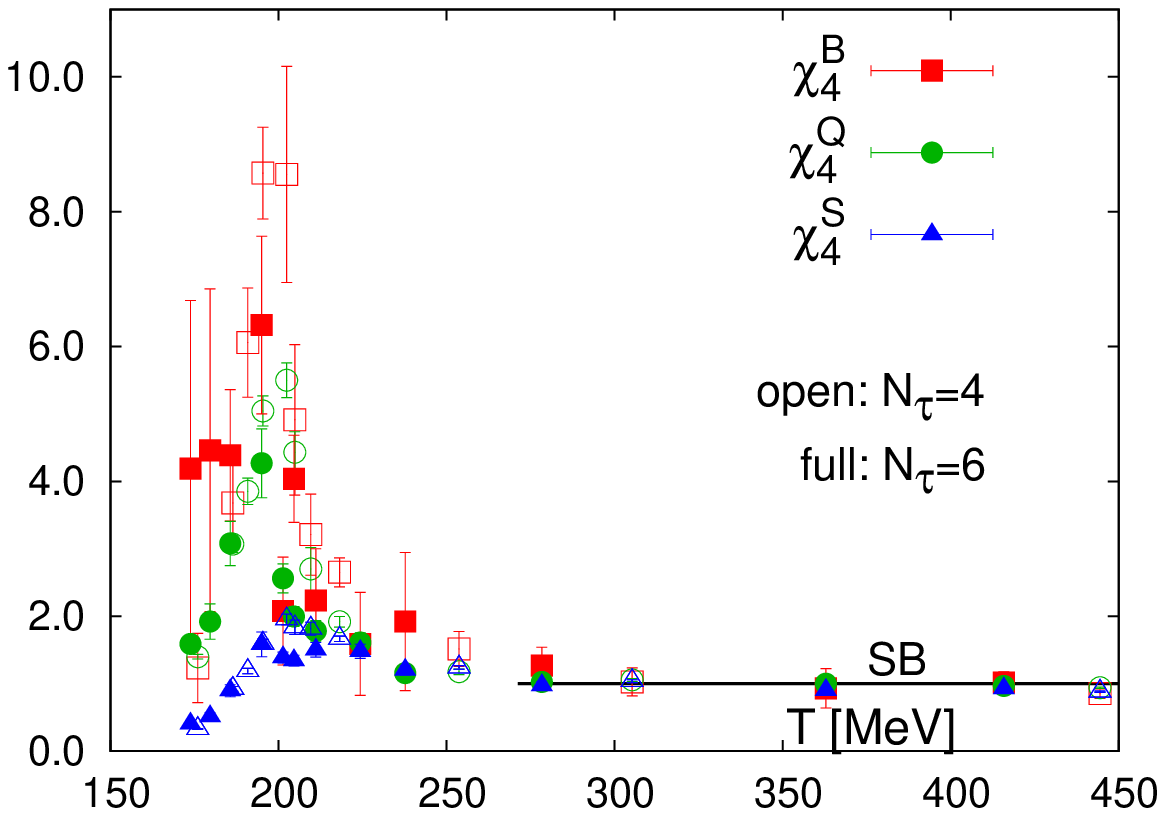}
\includegraphics[width=.32\textwidth]{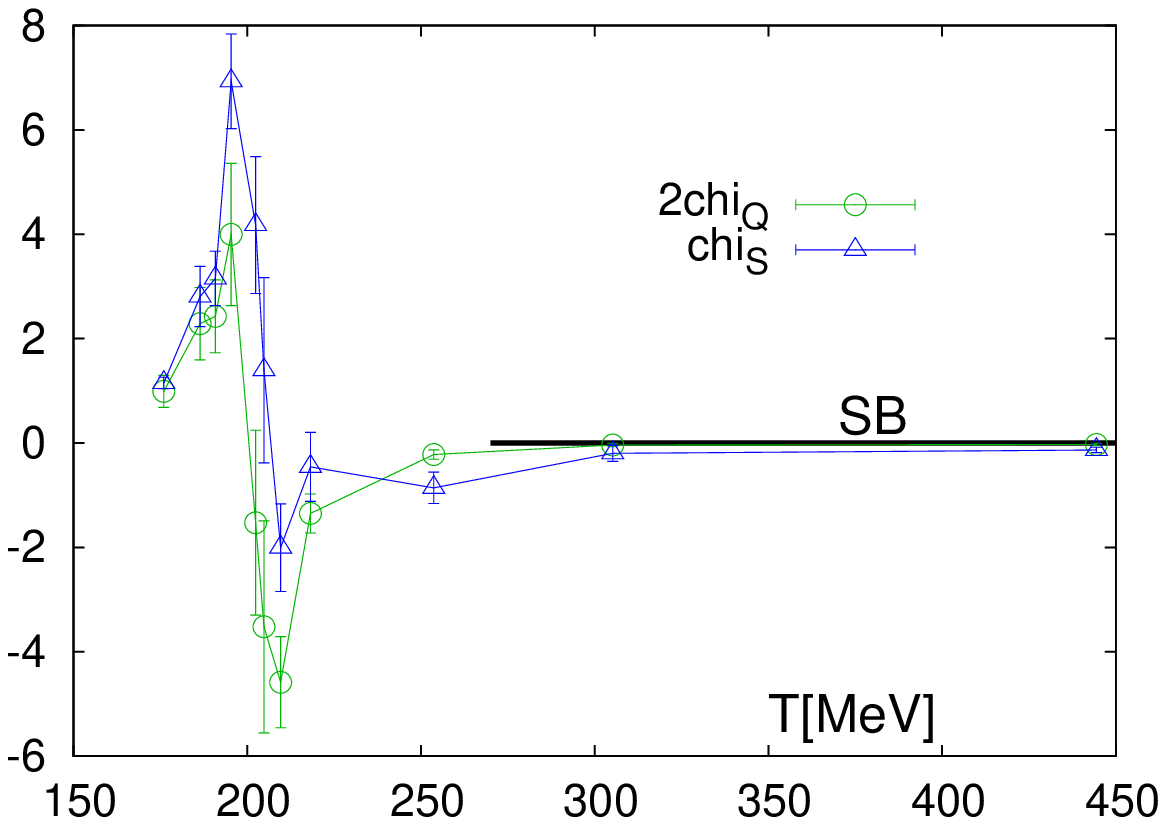}
\caption{Quadratic, quartic and 6th order cumulant of fluctuations for baryon number, electric charge and strangeness. All quantities have been normalized to the corresponding free quark gas values.}
\label{chi_BQS}
\end{figure}

In Fig. \ref{chi_BQS} we show results for quadratic fluctuations $\chi_2^X$, quartic fluctuations $\chi_4^X$ on $N_\tau=4$ and 6 lattices and 6th order cumulants of fluctuations $\chi_6^X$ on $N_\tau=4$ lattices, where $X=B,~Q$ and $S$. As can be seen, in all cases the quadratic fluctuations rise rapidly in the transition region where the quartic fluctuations show a maximum and 6th order cumulants change sign. This generic form is indeed expected. In the chiral limit the 4th order cumulants will have a peak and the 6th order cumulants will diverge, e.g. the Baryon number cumulants are expected to scale like
\begin{linenomath}
\begin{equation}
\chi_{2n}^B \sim \left| \frac{T-T_c}{T_c}\right|^{2-n-\alpha}
\end{equation}
\end{linenomath}
where $\alpha \simeq -0.25 [-0.015]$ denoting the critical exponent of the specific heat in 3-d, $O(4)$ [$O(2)$] spin models \cite{Onscaling}.

\begin{figure}[t]
\includegraphics[width=.24\textwidth]{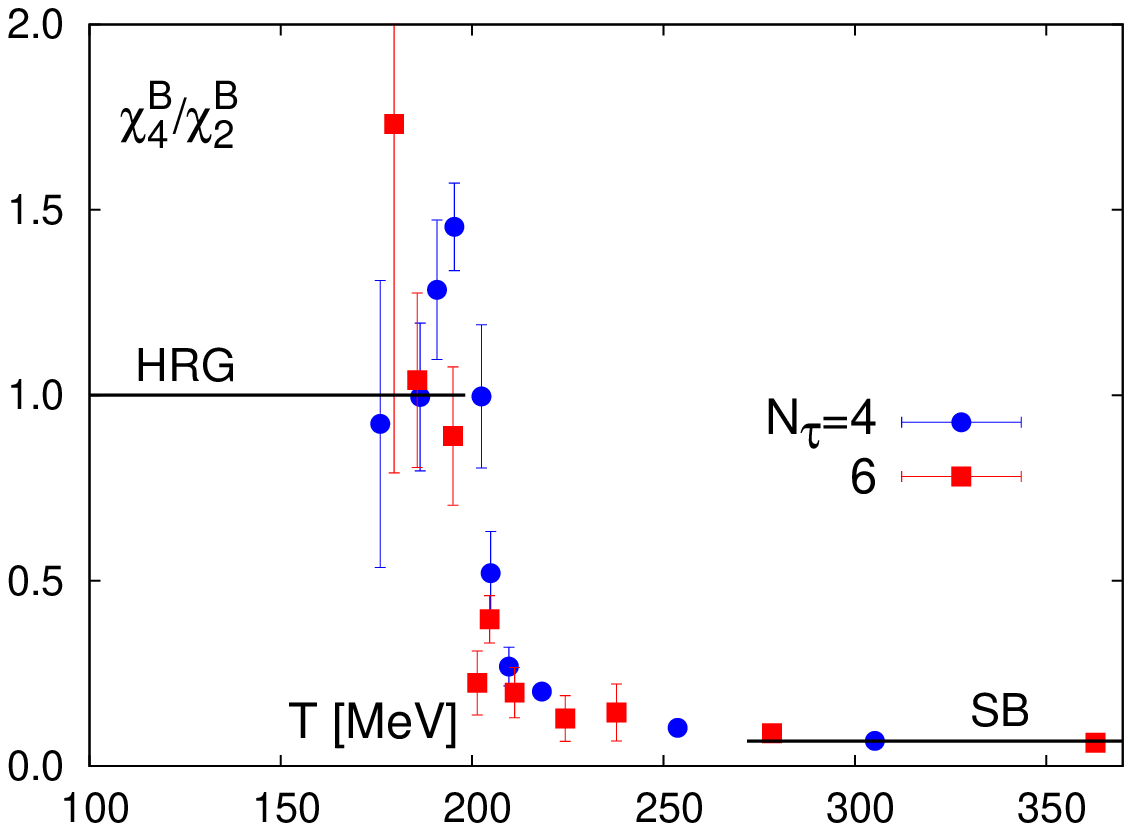}
\includegraphics[width=.24\textwidth]{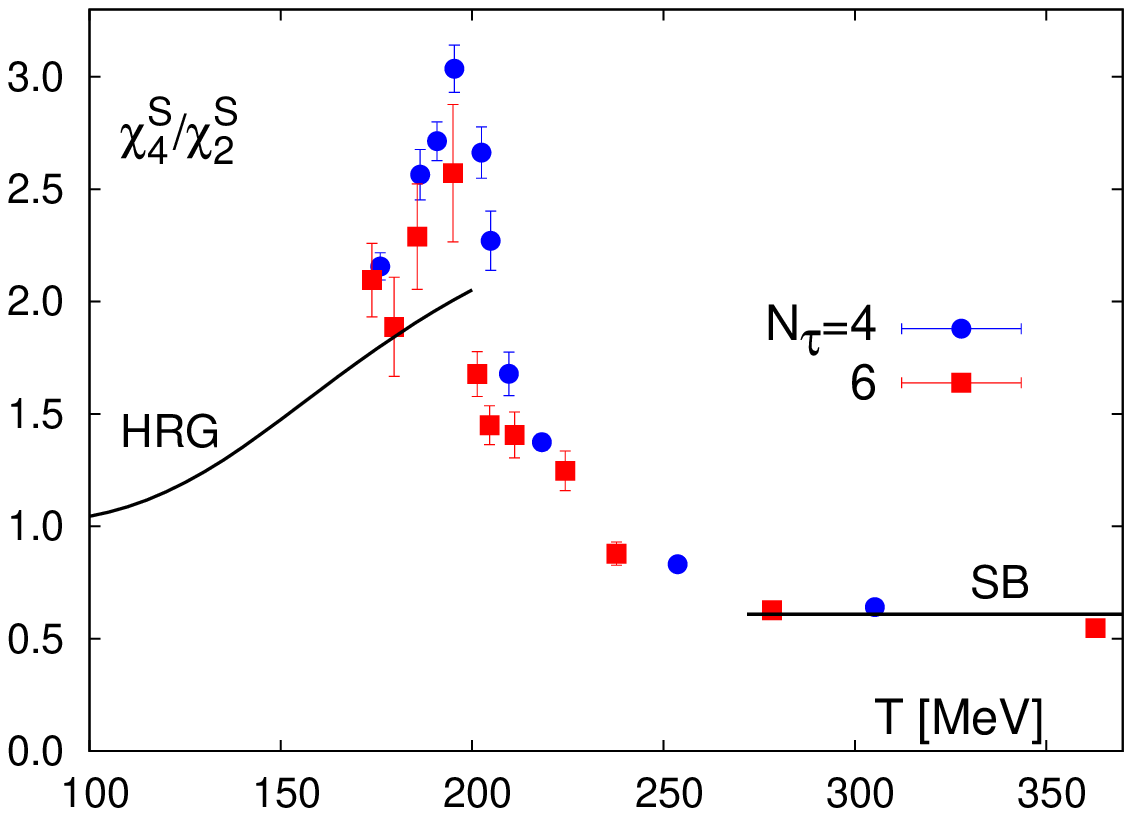}
\includegraphics[width=.24\textwidth]{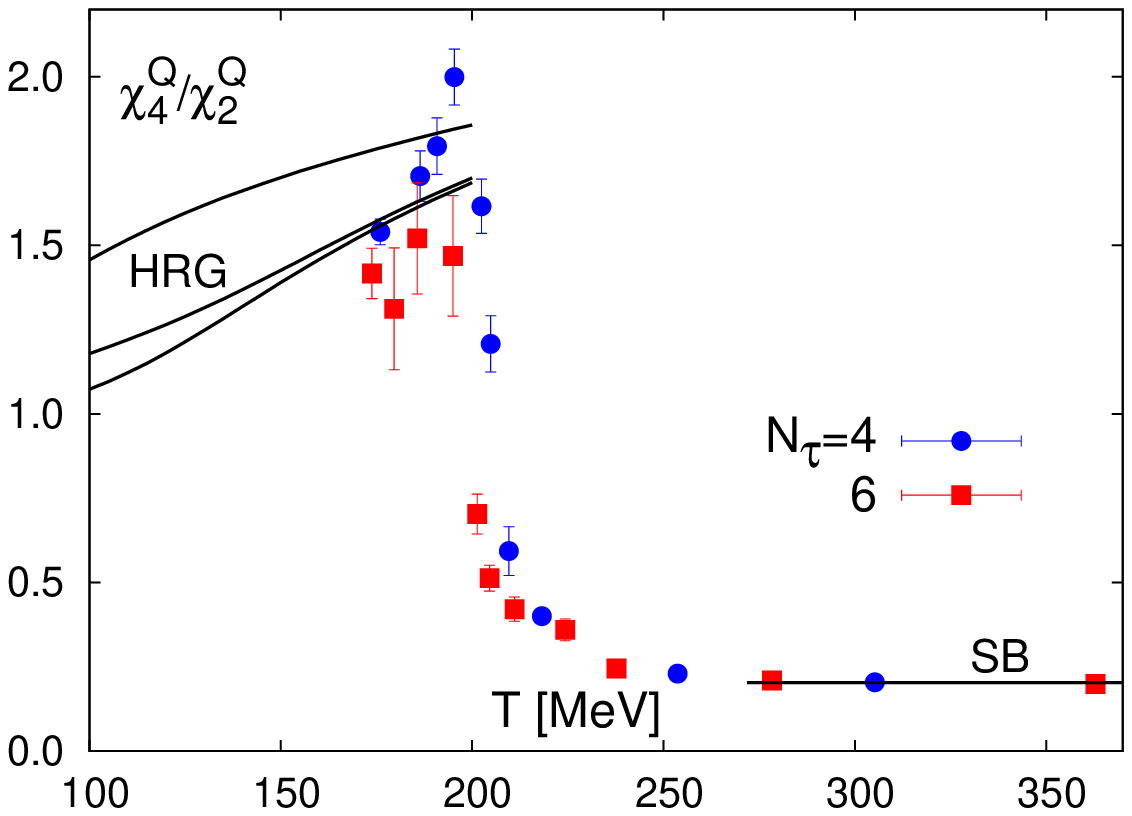}
\includegraphics[width=.24\textwidth]{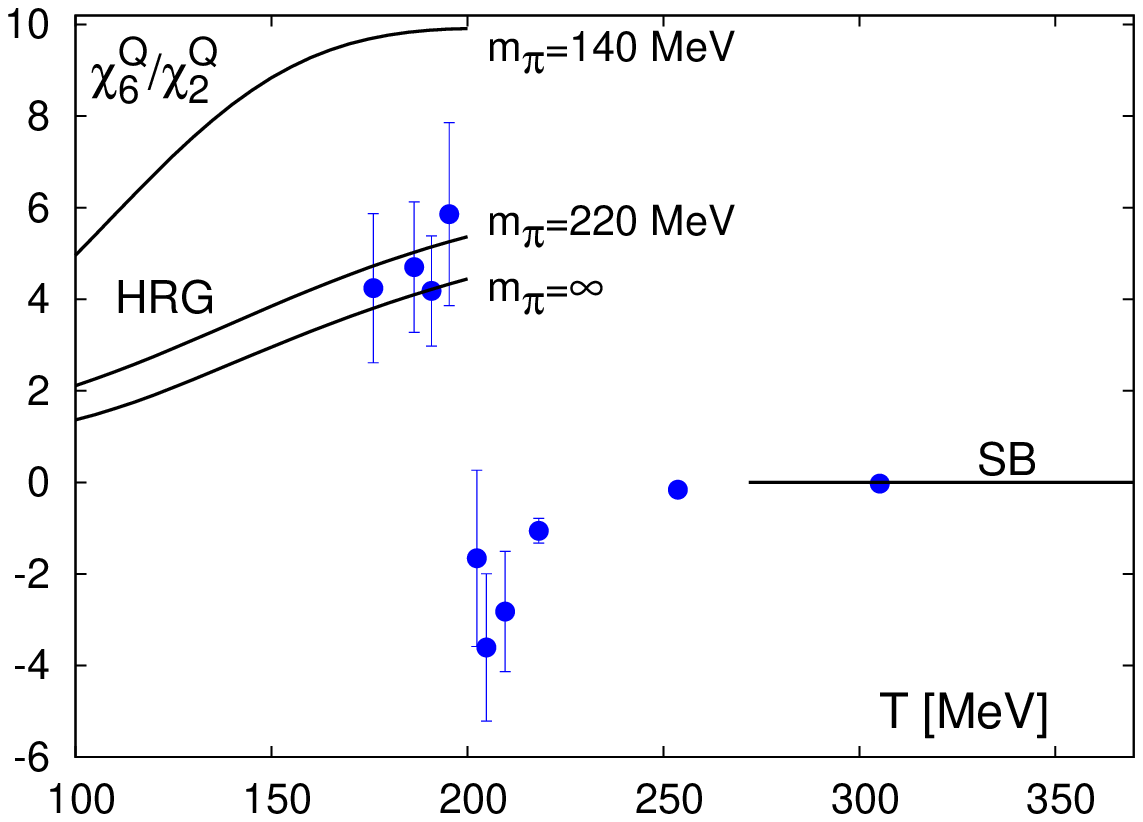}
\caption{The ratio of fourth and second order cumulants of baryon number (first), strangeness (second) and electric charge (third) fluctuations and the ratio of sixth and second order cumulants of electric charge fluctuations (fourth). In the last two cases we show curves for a HRG model calculated with physical pion masses (upper curve), pions of mass 220 MeV (middle) and infinitely heavy pions (lower curve).}
\label{ratio_cumu}
\end{figure}

At temperature  $T \gtrsim 1.5T_c$ the quadratic, quartic and 6th order cumulants of fluctuations of $B$, $Q$ and $S$ are well described by those of an ideal, massless quark gas. At low temperature, we compare lattice results for the ratios of the cumulants with HRG model calculations (Fig. \ref{ratio_cumu}). In the framework of a HRG model the ratio of fourth and second order cumulants of baryon number fluctuations is independent of the actual value of hadron masses; $(\chi_4^B/\chi_2^B)_{HRG}=1$ if all hadrons are heavy on the scale of the temperature. This is well reproduced by the lattice results. However, in the chiral limit it is expected that the cusp in $\chi_4^B$ (Fig. \ref{chi_BQS}) is expected to become more pronounced and thus more prominent also in the ratio $\chi_4^B/\chi_2^B$.

For $\chi_4^S/\chi_2^S$ and $\chi_4^Q/\chi_2^Q$, the lattice results also agree with HRG models qualitatively. The cumulants of electric charge fluctuations receive contribution from the lightest hadrons, pions. In our calculations the masses of the lightest pseudoscalar states, $220$~MeV, are close to but not exact as the physical pion masses. Therefore, we have analyzed the cumulant ratio $\chi_4^Q/\chi_2^Q$ in the HRG models with the pion masses at physical value $140$~MeV, $220$~MeV and infinity. We observe that the cumulant ratio is sensitive to the pion sector; with pion masses that are $50\%$ percent larger than the physical value, the contribution of the pion sector is drastically reduced. The pion sector contribution are more important for the 6th order cumulants than lower order cumulants. In Fig. \ref{ratio_cumu} (right) we show the cumulant ratio $\chi_6^Q/\chi_2^Q$. The lattice results agree with both HRG models without pion sector or with $220$~MeV pseudoscalar scalars, while the HRG with physical pions shows much larger discrepancy.
\section{Conclusion}
We have analyzed the fluctuations of baryon number, electric charge and strangeness in finite temperature QCD at vanishing chemical potential. We find fluctuations and correlations of conserved charges are well described by an ideal, massless quark gas already for temperatures of about ($1.5 \sim 1.7$) times the transition temperature. At low temperature we find that fluctuations of conserved charges are well described by a hadron resonance gas up to temperatures close to the
transition temperature.

The current analysis has been performed with light quarks that are one tenth of the strange quark mass. We have shown, that higher order cumulant ratios like, for instance $\chi_6^Q/\chi_2^Q$, become quite sensitive to the pseudoscalar Goldstone mass. In numerical calculations with staggered fermions this also means that results become sensitive to a correct representation of the entire Goldstone multiplet. Calculations with smaller quark masses closer to the continuum limit will thus be needed in the future to correctly resolve these higher order cumulants, which will give deeper insight into the range of applicability of the resonance gas model at low temperature and the non-perturbative features of the QGP above but close to Tc.




\end{document}